\documentclass[aps,prb,twocolumn,showpacs,superscriptaddress]{revtex4-1}

\usepackage{graphicx}
\usepackage{amsmath}

\begin{document}

\title{Low-temperature and dynamic magnetism of highly frustrated 5$d^2$ Li$_4$MgOsO$_6$ polymorphs in comparison with 5$d^3$ Li$_3$Mg$_2$OsO$_6$}

\author{Gia~T.~Tran}
\affiliation{Department of Chemistry and Biochemistry, California State University - Long Beach, Long Beach, CA 90840, USA}

\author{Phuong-Hieu~T.~Nguyen}
\affiliation{Department of Chemistry and Biochemistry, California State University - Long Beach, Long Beach,  CA 90840, USA}

\author{Charles~J.~Bloed}
\affiliation{Department of Chemistry and Biochemistry, California State University - Long Beach, Long Beach,  CA 90840, USA}

\author{Michael~E.~Evans}
\affiliation{Department of Physics, Villanova University, Villanova, PA 19085, USA}

\author{Jamie~A.~Anczarski}
\affiliation{Department of Physics, Villanova University, Villanova, PA 19085, USA}

\author{William~P.~Martin}
\affiliation{Department of Physics, Villanova University, Villanova, PA 19085, USA}

\author{Jefferson~Toro}
\affiliation{Department of Physics, Villanova University, Villanova, PA 19085, USA}

\author{Demetrios~V.~Papakostas}
\affiliation{Department of Physics, Villanova University, Villanova, PA 19085, USA}

\author{James~Beare}
\affiliation{Department of Physics and Astronomy, McMaster University, Hamilton, ON L8S 4M1 Canada}

\author{Murray~N.~Wilson}
\affiliation{Department of Physics and Astronomy, McMaster University, Hamilton, ON L8S 4M1 Canada}

\author{John~E.~Greedan}
\affiliation{Department of Chemistry and Chemical Biology, McMaster University, Hamilton, ON L8S 4M1 Canada}
\affiliation{Brockhouse Institute for Materials Research, McMaster University, Hamilton, ON L8S 4M1, Canada}

\author{Graeme~M.~Luke}
\affiliation{Department of Physics and Astronomy, McMaster University, Hamilton, ON L8S 4M1 Canada}
\affiliation{TRIUMF, 4004 Wesbrook Mall, Vancouver, BC V6T 2A3, Canada}

\author{Thomas~Gredig}
\affiliation{Department of Physics and Astronomy, California State University - Long Beach,  Long Beach, CA 90840, USA}

\author{Jeremy~P.~Carlo}
\altaffiliation[authors to whom correspondence should be addressed:  E-mail: jeremy.carlo@villanova.edu, shahab.derakhshan@csulb.edu]{}
\affiliation{Department of Physics, Villanova University, Villanova, PA 19085, USA}

\author{Shahab~Derakhshan}
\altaffiliation[authors to whom correspondence should be addressed:  E-mail: jeremy.carlo@villanova.edu, shahab.derakhshan@csulb.edu]{}
\affiliation{Department of Chemistry and Biochemistry, California State University - Long Beach, Long Beach,  CA 90840, USA}

\date{\today}

\begin{abstract}
Geometric magnetic frustration (GMF) has attracted substantial interest due to the exotic physics and rich phase diagrams revealed by 
the cancellation of normally-dominant magnetic interactions, giving impetus for the search for novel frustrated systems, most often based 
on antiferromagnetic correlations between magnetic ions decorating triangular or tetrahedral lattices.  We report here low-temperature 
magnetic susceptibility and muon spin relaxation results on Li$_4$MgOsO$_6$ and Li$_3$Mg$_2$OsO$_6$, members of the A$_5$BO$_6$ 
``rock salt ordered'' family of frustrated materials.  In Li$_3$Mg$_2$OsO$_6$ we find spin freezing below 12K.  In Li$_4$MgOsO$_6$, which 
can crystallize into either orthorhombic $Fddd$ or monoclinic $C2/m$ crystal symmetries depending on synthesis conditions, we find magnetism 
consistent with glassy-like behavior dominating below 2K, with partial ordering and evidence for dynamics at somewhat higher temperatures. 
\end{abstract}


\maketitle

\section{INTRODUCTION}

Antiferromagnetic (AFM) materials whose magnetic cations comprise triangular or tetrahedral sub-lattices are unable to satisfy spin correlation 
constraints simultaneously, resulting in a phenomenon known as geometric magnetic frustration (GMF).  In GMF systems the ground states are highly 
degenerate, which gives rise to rich magnetic phase diagrams exquisitely sensitive to external parameters that make them interesting topics among the 
condensed matter physics community.\cite{greedan, balents_nature} In systems exhibiting GMF the degree of frustration may be quantified by the frustration index, 
$f = \vert \Theta_W \vert / T_{N/f}$, where $\Theta_W$ is the Weiss temperature and $T_N$ or $T_f$ are either the N\'eel temperatures for long-range 
magnetic order or the spin freezing temperatures, respectively.\cite{schiffer}  

While this phenomenon has been extensively studied in transition metal oxides with the pyrochlore structure,\cite{gardner} more recently a great deal of attention has been 
devoted to systems with face-centered cubic (fcc) coordination of magnetic ions, namely ordered NaCl structure-type systems and B-site ordered 
double perovskites.\cite{howard,anderson_dblperovskites} 
Among the rock-salt type oxides, materials with the A$_5$BO$_6$ general formula have been the center of attention in our research program.  Here, A is a diamagnetic ion 
and B is a paramagnetic heavy ($4d$ or $5d$) transition metal ion. These systems are particularly interesting as the selected B ions may appear in various oxidation states, 
enabling a systematic study of the nature of magnetic ground state as a function of spin quantum numbers. In addition, high-$Z$ magnetic ions exhibit a moderate to high degree 
of spin-orbit coupling, which has been found to result in exotic physics in double perovskites \cite{thompson,russell,gangopadhyay,taylor,sarapulova,yuan,balents_new} and
other frustrated systems. Furthermore, these systems crystallize in several different crystal settings, which also provide benchmark examples for the study of structure-property 
relationships.  A$_5$BO$_6$ systems are most often found in two crystal systems, namely monoclinic ($C2/m$) and orthorhombic ($Fddd$). 

We have discovered and characterized several new members of this family with B = Ru, Re, and Os.  Li$_3$Mg$_2$RuO$_6$ \cite{derakhshan1} with Ru$^{5+}$ 
($S = 3/2$) ions was the first magnetic member of the family, which was shown to crystallize in the orthorhombic space 
group, $Fddd$. This compound undergoes a long-range order AFM transition at $\sim$17K, which was further confirmed by a lambda-shape anomaly in the heat capacity data as 
well as with temperature-dependent neutron diffraction data. It showed rather mild frustration with $f \sim 6$.  Subsequently, its osmate analog Li$_3$Mg$_2$OsO$_6$ was 
synthesized and studied.\cite{nguyen1} The latter isoelectronic and isostructural Os-base compound behaved strikingly differently from its ruthenate analogue. While a sharp peak 
in magnetic susceptibility data at $\sim 8K$ without any major divergence between zero-field-cooled (ZFC) and field-cooled (FC) conditions was indicative of long-range order, 
temperature-dependent heat capacity data revealed a very broad anomaly. This compound exhibits a relatively high frustration index of $\sim$13.  Furthermore, there were no 
magnetic neutron diffraction peaks down to 4K. Hence, the nature of the magnetic ground state of Li$_3$Mg$_2$OsO$_6$ has remained uncertain. 

Most recently, we were able 
to synthesize the $S = 1$ members of the family (Li$_4$MgOsO$_6$) in two different crystal settings, monoclinic $C2/m$ and orthorhombic $Fddd$.\cite{nguyen2}  The 
crystal structures of the two polymorphs are presented in Figures 1(a) and 1(b). The major difference between the arrangement of magnetic ions in these two systems lies in their 
dimensionality. While the orthorhombic lattice is composed of both 2D edge-sharing triangles and 3D structures resembling wedges (Figure 1(c)), the monoclinic phase comprises an 
edge-sharing triangular sub-lattice in a 2D fashion (Figure 2(d)). Nevertheless, the static magnetism in these two polymorphs were shown to be very similar. Both compounds 
showed no evidence of magnetic transition in temperature-dependent magnetic susceptibility data down to 2K.  The Curie-Weiss fits to the paramagnetic regime resulted in large, 
negative, and very similar Weiss constants ($-$115K and $-$122K for monoclinic and orthorhombic phases, respectively) indicating predominant AFM exchange correlations; the 
lack of a transition down to 2K indicates that both systems are highly frustrated ($f > 50$). 

However, it should be noted that $\Theta_W$ can be sensitive to the temperature range over which it is measured, and ordering temperature can be suppressed through mechanisms
other than frustration, thus $f$ can be influenced by factors besides geometric frustration.  Nonetheless, it is a useful rough metric for comparison of closely related specimens, 
although further study is required to confirm magnetic behavior.  Therefore, to better understand the ground state magnetic properties of these specimens, we report on low-temperature 
magnetic susceptibility and magnetic relaxation data for both phases of Li$_4$MgOsO$_6$ as well as $\mu$SR data for Li$_3$Mg$_2$OsO$_6$ and both Li$_4$MgOsO$_6$ phases.

 \begin{figure}[ht]
\includegraphics[width=90mm]{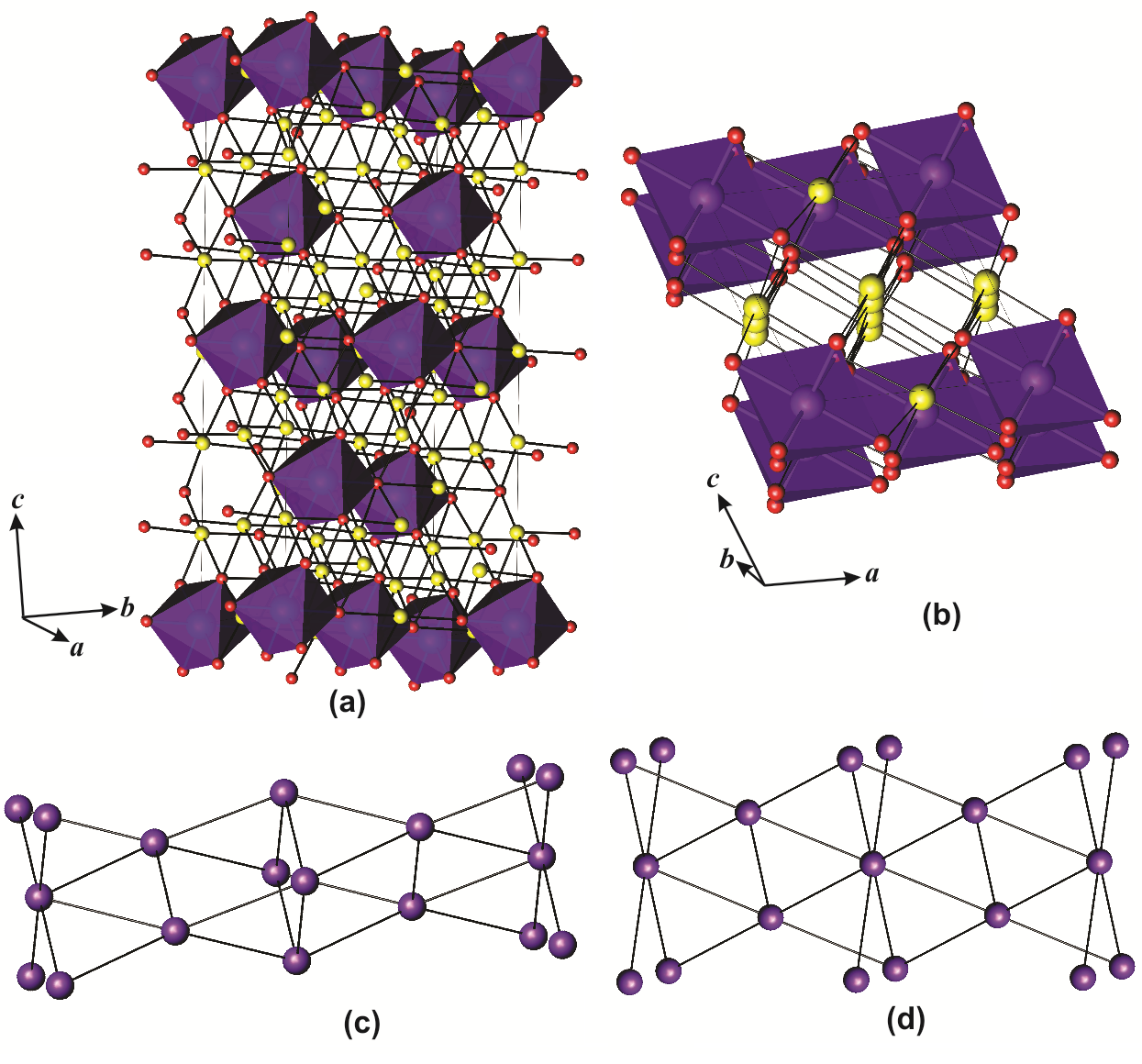}
\caption{\label{}(a-b) Crystal structures of (a) orthorhombic Li$_4$MgOsO$_6$, and (b) monoclinic Li$_4$MgOsO$_6$. The OsO$_6^{6-}$ octahedra are represented in purple, 
and the yellow spheres represent the diamagnetic ions (Li$^+$/Mg$^{2+}$). The red spheres are O$^{2-}$ ions.  (c-d) The magnetic sublattices of orthorhombic (c) and that of 
monoclinic Li$_4$MgOsO$_6$ (d). The purple spheres represent Os$^{6+}$ and the black lines indicate magnetic exchange pathways between the nearest neighbors.}
 \end{figure}

\section{EXPERIMENTAL}

\textit{Synthesis}. Nearly one gram samples of orthorhombic Li$_3$Mg$_2$OsO$_6$, orthorhombic Li$_4$MgOsO$_6$ and monoclinic Li$_4$MgOsO$_6$ were prepared 
by high-temperature solid state techniques. Details of these syntheses were explained in our previous reports.\cite{nguyen1,derakhshan2}  

\textit{Phase analyses}. To examine the formation and ensure the purity of produced phases, powder X-ray diffraction was employed. Data were collected using a 
PANalytical X$'$Pert Pro MPD diffractometer, equipped with a linear X$'$Celerator detector, with Cu-K$\alpha_1$ radiation. 

\textit{Magnetic susceptibility measurements}. ZFC and FC low-temperature susceptibility data were collected for both Li$_4$MgOsO$_6$ phases using a Quantum Design 
MPMS SQUID magnetometer equipped with an IQuantum $^3$He insert with a base temperature of 0.48K, under applied magnetic fields of 100 and 1000~Oe. 

\textit{Magnetization relaxation measurements}. Since the previously-studied static magnetic properties were almost identical for both Li$_4$MgOsO$_6$ phases,
 temperature-dependent magnetic relaxation data collections were performed with a Quantum Design PPMS vibrating sample magnetometer. For this 
purpose, samples were cooled down under zero field to the measurement temperature.  A magnetic field of 0.1 T was then applied and the time-dependent 
magnetization was measured for 5 to 30~ks at T = 5K, 10K, 20K, and 30K for both samples. 

\textit{Muon spin relaxation spectroscopy}.  To further characterize the magnetic ground state, muon spin relaxation ($\mu$SR) measurements were performed.  Muon spin relaxation is 
particularly useful in elucidating magnetic behavior in specimens with low-moment and/or spatially disordered magnetism, as are frequently encountered in geometrically frustrated 
systems, and can distinguish the effects of static order from dynamically fluctuating spins in systems exhibiting spin freezing.  In a $\mu$SR experiment, spin-polarized muons are 
implanted one at a time into a sample, within which each undergoes Larmor precession due to local magnetic fields at the implantation site.  The muons decay with a characteristic 
timescale of 2.2~$\mu$s, and emit positrons preferentially along the instantaneous spin axis of the muon at time of decay. The time and directional dependence of the positron emission 
can thus be used to reconstruct the time-dependent muon spin polarization function $G_z(t)$, from which the internal field distribution may be deduced.  The decay positrons are detected 
by a pair of counters on opposite sides of the sample, with the Asymmetry defined as the difference between the count rates in the two detectors divided by the sum.  Note that due to 
differing detector efficiencies and geometry, the baseline raw asymmetry (corresponding to zero net muon spin polarization) can be different from zero. Measurements were conducted at 
TRIUMF (Vancouver, BC) using the M20 beamline with 4.2 MeV surface muons and the LAMPF spectrometer at temperatures from 2 to 125K, in both zero field (ZF) and longitudinal field (LF) 
configurations.  

\section{RESULTS / DISCUSSION}

Low temperature ZFC and FC SQUID data are shown for the monoclinic and orthorhombic Li$_4$MgOsO$_6$ phases in Figures 2(a) and 2(b), respectively. Clear AFM-type transitions are observed 
for both $C2/m$ and $Fddd$  phases at 1.5K and 1.3K, respectively, corresponding to frustration indices $f$ = 77 and 94. Such divergence is indicative of spin-glass 
behavior for both samples, with a slightly lower transition temperature in the orthorhombic phase as compared to the monoclinic. The spin-glass magnetic ground state is also
consistent with the cationic occupancy disorder between Li and Mg ions in crystallographic cationic position.\cite{zvereva}

\begin{figure}[ht]
 \includegraphics[width=85mm]{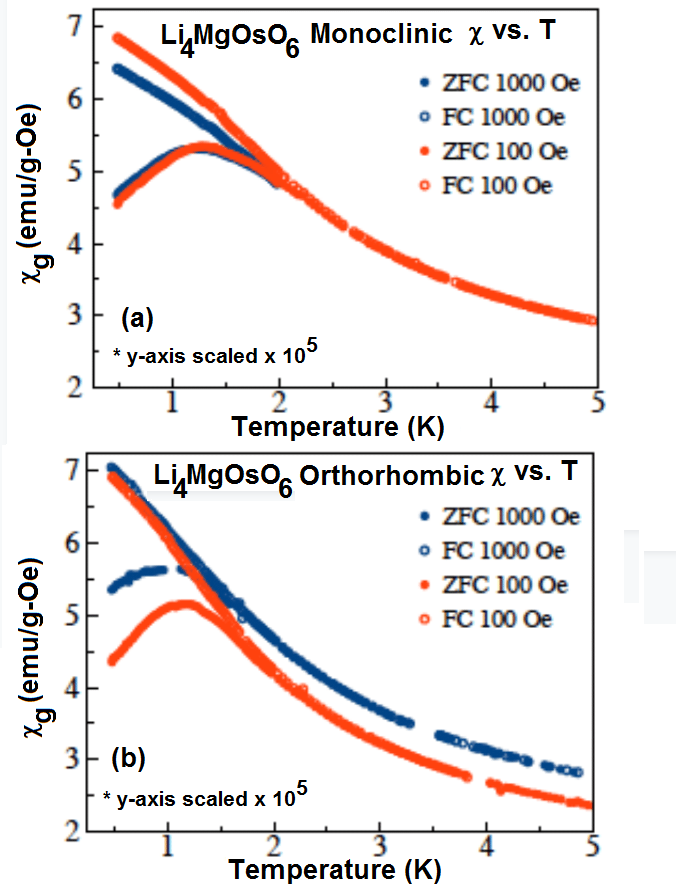}
 \caption{\label{}Temperature-dependent zero-field cooled/field cooled (ZFC, FC) magnetic susceptibility data for (a) monoclinic Li$_4$MgOsO$_6$ and (b) orthorhombic 
Li$_4$MgOsO$_6$ under applied magnetic fields of 100 Oe (red) and 1000 Oe (blue).}
 \end{figure}

Both phases of Li$_4$MgOsO$_6$ show slow magnetic relaxation, albeit with diﬀerent long-term behavior, as shown in Figure 3. The magnetization is measured after 
a magnetic field of 0.1~T is established and increases in magnitude for lower temperatures. For easier comparison, the magnetization $M(t)$ is normalized to the
 asymptotic magnetization value $M_\infty$, which is obtained from a single exponential fit to the experimental data. In the case of the monoclinic phase, the 
magnetic relaxation is more pronounced. While at 5, 10 and 20K the magnetization is still increasing over time, at 30K the opposite trend is observed. At the 
lowest temperatures, the sweep field speed of 100~Oe per second is faster than the sample response time, so that the magnetization is increasing in response 
to the applied field, whereas it is likely an aging effect gives rise to the magnetization decrease at higher temepratures. All data can be fit to a single exponential with 
a typical time constant of around 3000~s. These fits are shown as solid lines in Figure 3. In the case of the orthorhombic phase, there is a decrease for all 
temperatures, except at the lowest measured temperature of 5K, in which case a double exponential is used to capture both the initial increase that is then 
followed by the relaxation of the magnetization. This slow magnetic relaxation is characteristic of glassy systems.

 \begin{figure}[h]
 \includegraphics[width=85mm]{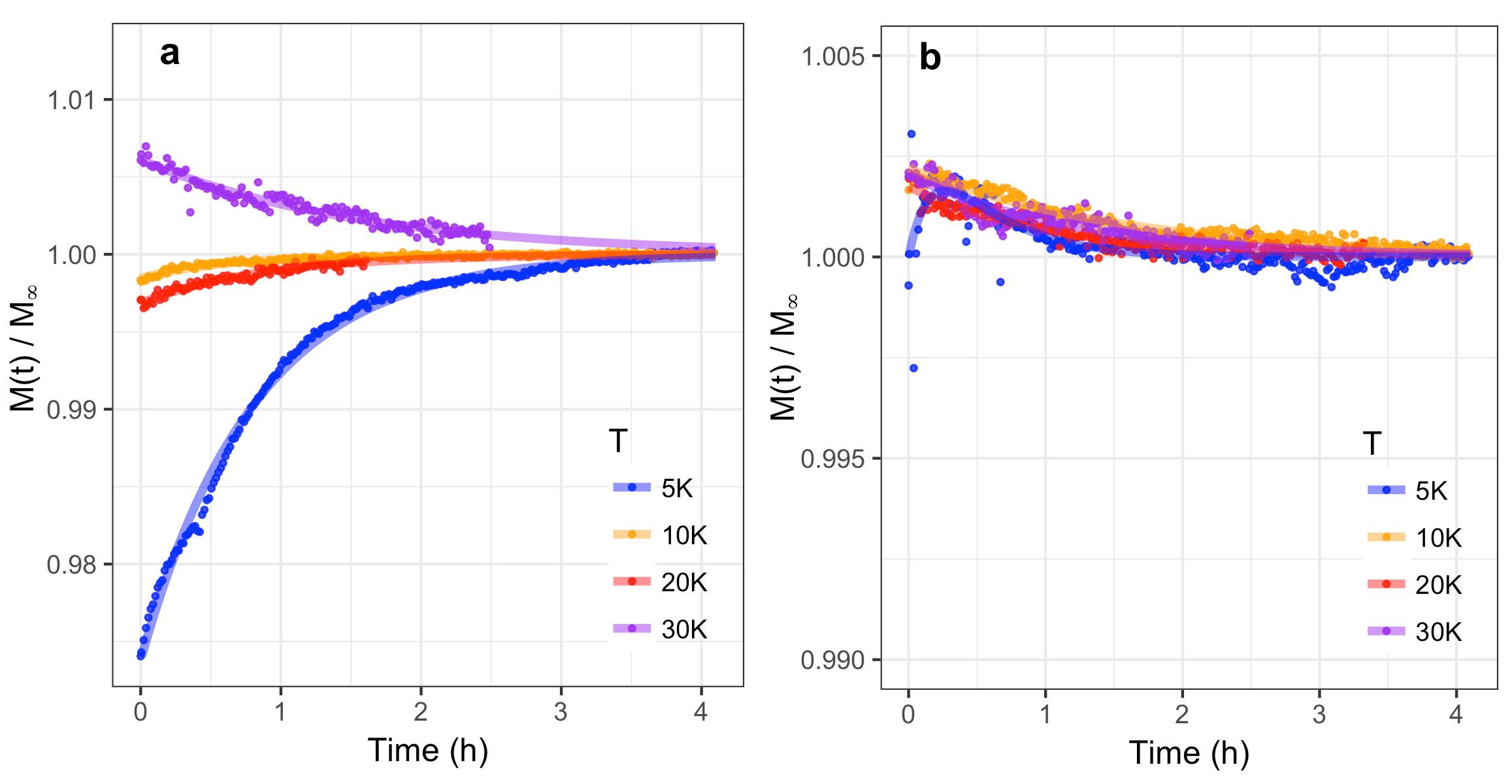}
 \caption{\label{}The time-dependent magnetization $M(t)$ measured at four different temperatures in an applied field of 0.1~T for (a) the monoclinic phase, and (b) the
 orthorhombic phase, using a vibrating sample magnetometer. The data is normalized to the asymptotic magnetization value $M_\infty$ for comparison at different temperatures. 
The solid lines are fits to a single exponential function, and a double exponential for $T$=5K. }
\end{figure}

The zero-field $\mu$SR data for Li$_3$Mg$_2$OsO$_6$ are shown in Figure 4(a), evidencing an onset of relaxation below 12K, consistent with glassy-type ordering.  
These data were fitted to the Uemura spin glass function\cite{uemura} which was developed for dilute magnetic alloys but has been used more generally for frustrated 
small-moment systems exhibiting glassy or spatially complex magnetism: 

\begin{equation}	 
\begin{split}
G_z(t) = \frac{1}{3}exp\left(-\sqrt{\lambda_dt}\right)\quad +  \quad \quad  \quad\\
\frac{2}{3}\left(1-\frac{a_s^2t^2}{\sqrt{\lambda_dt + a_s^2t^2}}\right) exp\left(-\sqrt{\lambda_dt+a_s^2t^2}\right)  
\end{split}
\end{equation}

Here, $a_s$ represents the statically ordered moment, while $\lambda_d$ represents relaxation due to dynamically fluctuating moments.   The 2/3 term corresponds to the 
component of magnetic moments perpendicular to the initial muon spin asymmetry, which are relaxed by both the static and dynamic components of magnetism, whereas the 1/3 
term represents the component parallel to the initial muon spin asymmetry, which is only relaxed by dynamically fluctuating moments. Fits to the Uemura spin glass function 
are shown in Figure 4b, exhibiting an onset of static relaxation starting just above 12K, with dynamical fluctuations peaking somewhat below 10K.  The rise in $a_s$ and
the peak in $\lambda_d$ are 
consistent with a progressive slowing down of spin fluctuations, resulting in static order at base temperature.  Since the 1/3 and 2/3 terms have similar functional forms 
and can exhibit significant interplay in fitting, a total relaxation equal to $\sqrt{\lambda_d^2+ a_s^2}$ is also shown and exhibits an order-parameter-like dependence with 
onset at 12K with the relaxation at base temperature approaching 31 $\mu$$s^{-1}$.   It should be noted the slow relaxation evident at higher temperatures (e.g. 25K) is due to 
nuclear dipolar relaxation from the large Li nuclear moments. 

\begin{figure}[h]
\includegraphics[width=85mm]{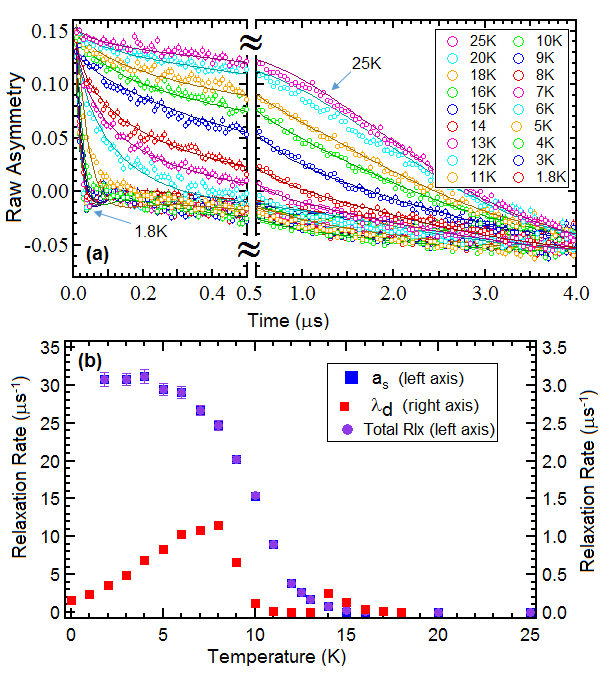}
\caption{\label{}(a) Muon spin relaxation data for Li$_3$Mg$_2$OsO$_6$, fitted to the Uemura spin glass function as described in the text.  The short-time and 
long-time domains are shown with separate scaling to emphasize both the rapid initial relaxation and the slower long-time behavior.   For clarity, the highest
and lowest temperature traces are individually labeled.  (b) Fits to the Uemura spin glass function of muon spin relaxation data for Li$_3$Mg$_2$OsO$_6$, 
exhibiting an onset of magnetic order below 12K. For clarity, $a_s$ and total relaxation $\sqrt{a_s^2+\lambda_d^2}$ are scaled by the left y-axis, while $\lambda_d$ uses the right y-axis. }
\end{figure}

$\mu$SR data for both crystal settings of Li$_4$MgOsO$_6$ are shown in Figures 5(a) and 5(b).  Both specimens exhibit onset of low-temperature relaxation, 
albeit at a temperature scale approximately half as high as in Li$_3$Mg$_2$OsO$_6$, and with low-temperature relaxation rates about twenty times smaller, 
corresponding to a commensurately smaller ordered moment size.  These data were also fitted to the Uemura spin-glass function, as shown in Fig. 5(c-d).   In both 
cases the static moment $a_s$ exhibits order-parameter-like dependence commencing at around 5K, while dynamical relaxation $\lambda_d$ exhibits a peak and then
declines toward zero as the fluctuations slow down through the muon's characteristic time window.

\begin{figure}[h]
\includegraphics[width=85mm]{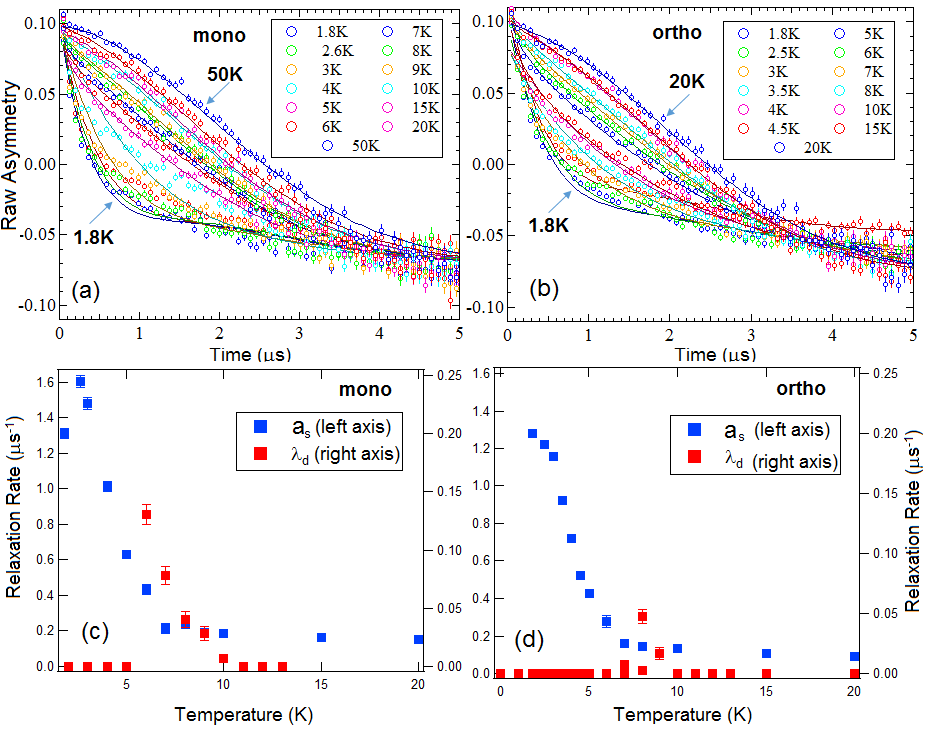}
\caption{\label{}(a-b) Muon spin relaxation data for Li$_4$MgOsO$_6$, fitted to the Uemura spin glass function as described in the text; the monoclinic specimen is shown 
at left (a), with the orthorhombic specimen at right (b).  For clarity, the highest and lowest temperature traces are labeled individually.  (c-d) Fits to the Uemura spin-glass 
function for monoclinic (c) and orthorhombic (d) Li$_4$MgOsO$_6$.  Both exhibit onset of relaxation just above 5K, and become fully ordered at a temperature below 2K. }
\end{figure}

In both specimens, relaxation becomes apparent at around 5K, although full order does not appear to set in until 2K or below, consistent with the low-temperature 
susceptibility measurements.   As with Li$_3$Mg$_2$OsO$_6$, the high-temperature relaxation is due to Li nuclear dipolar moments.  The relaxation rates at base 
temperature are 1.6~$\mu$$s^{-1}$ in the monoclinic specimen and 1.3~$\mu$$s^{-1}$ in the orthorhombic setting.  This corresponds to a 20-25\% larger ordered 
moment size in the monoclinic specimen, commensurate with the observed ratio of ordering temperatures.

\section{CONCLUSIONS}

We have studied the rock-salt ordered antiferromagnets Li$_3$Mg$_2$OsO$_6$ and Li$_4$MgOsO$_6$ in the context of geometric magnetic frustration.  These
systems most often crystallize into either orthorhombic $Fddd$ or monoclinic $C2/m$ crystal settings; Li$_4$MgOsO$_6$ in particular can crystallize into either
of these crystal settings depending on synthesis conditions, making its study especially valuable in elucidating ground state determination in rock-salt ordered
systems.  Li$_3$Mg$_2$OsO$_6$ and both crystal settings of Li$_4$MgOsO$_6$ are observed to exhibit magnetic ordering consistent with spin freezing, with 
dynamical slowing down of fluctuations observed while crossing through the transition.  In Li$_3$Mg$_2$OsO$_6$, the spin-freezing temperature indicated by 
muon spin relaxation is 12K. While in both specimens of Li$_4$MgOsO$_6$ the onset of relaxation in $\mu$SR data occurs around 5K, low-temperature magnetic 
susceptibility measurements indicate an ordering temperature in both of 1.5K in the monoclinic specimen and 1.3K in the orthorhombic specimen, corresponding to 
high frustration indices of 77 and 94, respectively.  The ordered moment size and the ordering temperature in monoclinic Li$_4$MgOsO$_6$ is about 20\% larger 
than in its orthorhombic polymorph, highlighting the significance to ground state determination of subtle structural distortions and differing magnetic pathways, as
depicted in Figure 1.  The ordered moment size in the $5d^3$ Li$_3$Mg$_2$OsO$_6$ is about 20 times larger than the moment size in either specimen of $5d^2$ 
Li$_4$MgOsO$_6$, although the fully ordered moment size may not be achieved by the lowest temperature (1.8K) accessible to the muon spin relaxation 
experiments.   

\begin{acknowledgments}
SD is grateful for financial support from NSF-DMR-RUI Award \#1601811. SD and TG also acknowledge support from W.~M.~Keck Foundation for establishment of the 
Keck Energy Materials Program at CSULB. JPC acknowledges support from the Research Corporation for Science Advancement (Cottrell College Science Award \#23314).  
We thank the TRIUMF CMMS for assistance with $\mu$SR experiments. 
\end{acknowledgments}


%

\end{document}